\documentclass[aps,pra,twocolumn,superscriptaddress]{revtex4-2}
\usepackage{graphicx}
\graphicspath{{../},{figures/}}
\usepackage{amssymb}
\usepackage{amsmath}
\usepackage[utf8x]{inputenc}

\usepackage{url}
\usepackage[colorlinks]{hyperref}
\hypersetup{%
    plainpages=true,
    breaklinks=true,
    hypertexnames=false,
    pageanchor=true,
    colorlinks=true,
    linkcolor={blue},
    citecolor={red},
    urlcolor={blue},
    anchorcolor={black}
    }

\newcommand{\ket}[1]{| #1 \rangle}
\newcommand{\bra}[1]{\langle #1 |}

\def\tr{{\rm Tr}}

\def\FEF{{\rm FEF}}

\def\<{\langle}
\def\>{\rangle}

\newcommand{\bs}{|\psi^-\rangle}

\begin{document}


\title{Experimental hierarchy of two-qubit quantum correlations without state
tomography}

\author{Shilan Abo} 
\affiliation{Institute of Spintronics and Quantum Information,
Faculty of Physics, Adam Mickiewicz University, 61-614 Pozna\'n,
Poland}

\author{Jan Soubusta} \email{jan.soubusta@upol.cz}
\affiliation{Palack\'y University Olomouc, Faculty of Science,
Joint Laboratory of Optics of PU and IP CAS,
17. listopadu 1192/12, 779 00 Olomouc, Czech Republic}

\author{Kate\v{r}ina Jir\'akov\'a} 
\affiliation{Palack\'y University Olomouc, Faculty of Science,
Joint Laboratory of Optics of PU and IP CAS,
17. listopadu 1192/12, 779 00 Olomouc, Czech Republic}

\author{Karol Bartkiewicz} 
\affiliation{Palack\'y University Olomouc, Faculty of Science,
Joint Laboratory of Optics of PU and IP CAS,
17. listopadu 1192/12, 779 00 Olomouc, Czech Republic}
\affiliation{Institute of Spintronics and Quantum Information,
Faculty of Physics, Adam Mickiewicz University, 61-614 Pozna\'n,
Poland}

\author{Anton\'in \v{C}ernoch} 
\affiliation{Institute of Physics of the Czech Academy of
Sciences, Joint Laboratory of Optics of PU and IP CAS, 17.
listopadu 1154/50a, 779 00 Olomouc, Czech Republic}

\author{Karel Lemr} \email{k.lemr@upol.cz}
\affiliation{Palack\'y University Olomouc, Faculty of Science,
Joint Laboratory of Optics of PU and IP CAS,
17. listopadu 1192/12, 779 00 Olomouc, Czech Republic}

\author{Adam Miranowicz} \email{miran@amu.edu.pl}
\affiliation{Institute of Spintronics and Quantum Information,
Faculty of Physics, Adam Mickiewicz University, 61-614 Pozna\'n,
Poland}

\begin{abstract}
A Werner state, which is the singlet Bell state affected by white
noise, is a prototype example of states, which can
reveal a hierarchy of quantum entanglement, steering, and Bell
nonlocality by controlling the amount of noise. However,
experimental demonstrations of this hierarchy in a sufficient and
necessary way (i.e., by applying measures or universal witnesses
of these quantum correlations) have been mainly based on full quantum state
tomography, corresponding to measuring at least 15 real parameters of
two-qubit states. Here we report an experimental demonstration of
this hierarchy by measuring only six elements of a correlation
matrix depending on linear combinations of two-qubit Stokes
parameters. We show that our experimental setup can also reveal
the hierarchy of these quantum correlations of generalized Werner
states, which are any two-qubit pure states affected by white
noise.
\end{abstract}

\date{\today}

\maketitle

\section*{Introduction}

Quantum correlations reveal not only the strangeness of quantum
mechanics, but are the main resources for quantum technologies,
including quantum sensing and quantum information
processing~\cite{HidaryBook}. Thus, the detection, control, and
quantification of these resources is of paramount importance.

Among different types of correlations, a special interest has been
paid to quantum entanglement~\cite{HorodeckiReview},
Einstein-Podolsky-Rosen (EPR) steering (also called quantum
steering)~\cite{CavalcantiReview, UolaReview}, and Bell
nonlocality that can be revealed by testing the violation of a
Bell inequality~\cite{BrunnerReview}. These types of correlations
coincide for two-qubit pure states, but can be different for mixed
states. Probably, the most intuitive distinction between these
three types of quantum correlations for two systems (parties) can
be given from a cryptographic perspective with the use of trusted
and untrusted detectors. Specifically, according to
Refs.~\cite{Wiseman2007,Jones2007}: (i) quantum entanglement can
be revealed if both parties use only trusted detectors, (ii) EPR
steering can be tested if one party uses trusted detectors and the
other untrusted ones, and (iii) Bell nonlocality can be
demonstrated if both parties use untrusted detectors. We
experimentally determined and compared measures of these
correlations for Werner states.

It is theoretically well known that by gradually adding noise to a
pure state, one can reveal a hierarchy of different types of
quantum correlations, including quantum entanglement, EPR
steering, and Bell nonlocality. These effects are equivalent for
two-qubit pure states, however they are in general different for
mixed states. Werner~\cite{Werner1989} reported in 1989 that the
singlet Bell state affected by white noise can be entangled
without exhibiting Bell nonlocality, i.e., without violating any
Bell inequality. It was further found that the Werner state with a
proper amount of white noise can be entangled but unsteerable, or
steerable without exhibiting Bell nonlocality, in addition to the
trivial cases when a given state is Bell nonlocal (so also
steerable and entangled) or separable (so also unsteerable and
Bell local). Even a more refined hierarchy can be revealed by
considering generalized Werner states, defined as mixtures of an
arbitrary two-qubit pure states and white
noise~\cite{Jirakova2021}. Thus, the Werner and Werner-like states
can be considered prototype examples of states indicating such a
hierarchy. Their generation and the detection of their quantum
correlations are a central topic of this paper.

Here we study a hierarchy of quantum correlations via their
measures. We note that various other hierarchies of non-universal
witnesses of quantum correlations have been investigated in
detail. These include studies of sufficient conditions (i.e.,
nonuniversal witnesses) for observing specific types of
correlations via matrices of moments of, e.g., the annihilation,
creation, position, momentum, or Pauli operators. For example: (i)
hierarchies of various witnesses of spatial~\cite{Richter2002} and
spatiotemporal~\cite{Vogel2008,Miranowicz2010} correlations of
bosonic systems revealing their nonclassicality via a nonpositive
Glauber-Sudarshan $P$ function; (ii) a hierarchy of entanglement
witnesses~\cite{Shchukin2005,Miranowicz2006} based on the
Peres-Horodecki partial transposition criterion or their
generalizations~\cite{Miranowicz2009} using contraction maps
(e.g., realignment) and positive maps (e.g., those of Kossakowski,
Choi, and Breuer); (iii) a hierarchy of necessary conditions for
the correlations that arise when performing local measurements on
separate quantum systems, which enabled finding a hierarchy of
upper bounds on Bell nonlocality~\cite{Navascues2007,
Navascues2008} (iv) a hierarchy of EPR steering
witnesses~\cite{Kogias2015} based on entanglement criteria with
the constraint that measurement devices of one party cannot be
trusted. Especially powerful methods for finding infinite
hierarchies of quantum-correlation criteria are those formulated
as semidefinite programs~\cite{Doherty2004, Eisert2004,
Navascues2007, Navascues2008, Kogias2015}. Note that semidefinite
programming has been found very effective in calculating not only
nonuniversal witnesses but also measures (or universal witnesses)
of quantum steering~\cite{Skrzypczyk2014, Piani2015,
CavalcantiReview}, Bell nonlocality~\cite{BrunnerReview}, and
entanglement~\cite{HorodeckiReview}. It could also be noted that a
hierarchy of quantum nonbreaking channels, which is closely
related to a hierarchy of temporal~\cite{Ku2018hierarchy} and
spatial quantum correlations, has been studied very recently both
theoretically and experimentally in Ref.~\cite{Ku2022}, where the
effects of white noise (or, equivalently, a qubit-depolarizing
channel) on quantum memory, temporal steerability~\cite{Chen2016,
Chen2017}, and nonmacrorealism were revealed by applying a
\emph{full} quantum process tomography.

The use of measures or universal witnesses of these quantum
correlations, however, is required to demonstrate experimentally
such hierarchies in a sufficient and necessary manner. For
example, to our knowledge, no experiment has been performed to
determine standard entanglement measures of a general two-qubit
mixed state without full quantum state tomography (QST). These
measures include the concurrence~\cite{Hill1997}, which is a
measure of the entanglement of formation, the
negativity~\cite{Zyczkowski1998} quantifying the Peres-Horodecki
entanglement criterion, and the relative entropy of
entanglement~\cite{Vedral1998}. Thus, a QST-based approach to
study a hierarchy of quantum correlations was applied in our
former related study~\cite{Jirakova2021}, as based on measuring 16
real parameters for two-qubit Werner states.

A hierarchy of quantum-correlation measures enables efficient
estimations of one measure for a given value of another. More
specifically, estimating the range of a measure of a given type of
quantum correlation for a certain value of a measure (or bounds)
of another type of quantum correlations were reported for
arbitrary or specific classes of two-qubit states. These
estimations include comparisons of: (i) entanglement and Bell
nonlocality~\cite{Verstraete2002,
Bartkiewicz2013,Horst2013,Su2020}, (ii) steering and Bell
nonlocality~\cite{Quan2017}, as well as (iii) entanglement and
steering~\cite{Fan2021}. Note that such estimations can also be
applied to compare non-equivalent measures describing the same
type of correlations, including two-qubit
entanglement~\cite{Adam2004a, *Adam2004b} or single-qubit
nonclassicality~\cite{Adam2015}. Explorations of the relationships
between measures of entanglement, steering, and Bell nonlocality
for specific types of two-qubit states have been also attracting a
considerable interest. Recent studies include, e.g., theoretical
analyses of two-qubit $X$-states~\cite{Sun2017} and two-mode
Gaussian states~\cite{Qureshi2018}. Experimental QST-based
hierarchies of quantum entanglement, steering, and Bell
nonlocality for specific classes of two-qubit states in relation
to the above-mentioned estimations were also reported. These
include experiments with mixtures of partially entangled two-qubit
pure states~\cite{Yang2021} and GWSs based on full
QST~\cite{Jirakova2021} or full quantum process
tomography~\cite{Ku2022}. A hierarchy of the latter states is also
experimentally studied here but \emph{without} applying a full
QST.

We note that an experimental method for testing polarization
entanglement without QST of general two qubit states was proposed
in Ref.~\cite{Bartkiewicz2015} based on measuring a collective
universal witness of Ref.~\cite{Augusiak2008}. However, the
method, to our knowledge, has not been implemented experimentally
yet. Another experimental approach to determine entanglement of a
given state without QST can be based on measuring a bipartite
Schmidt number, which satisfies various conditions of a good
entanglement measure~\cite{Terhal2000,Sanpera2001} and can be
determined experimentally via a witnessing
approach~\cite{Sperling2011}. However, it is not clear how the
same method can also be used to experimentally determine also
steering and nonlocality measures. Note that we want to apply a
versatile experimental setup, which can be used to determine
various measures of all the three types of quantum correlations.

Multiple indicators of quantum steering have been demonstrated
experimentally (for a review see Ref.~\cite{UolaReview}). We note
a very recent Ref.~\cite{Fan2023}, where it was shown
experimentally that a critical steering radius is the most
powerful among practical steering indicators. Its scaling property
allows classifying as steerable or non-steerable various families
of quantum states. This approach is useful in testing theoretical
concepts of the critical radius in real experiments prone to
unavoidable noise. The authors used a setup introducing losses and
measured elements of a correlation matrix to determine the
steering indicators. Similar quantifiers, but describing
nonlocality and entanglement, were measured in
Ref.~\cite{Bartkiewicz2017} using the parameters $M$ and $F$,
which are also applied in this paper.

Here, we report the first (to our knowledge) experimental
demonstration of the hierarchy of measures of entanglement,
steering, and Bell nonlocality without applying full QST, i.e., by
measuring only six elements of a correlation matrix $R$
(corresponding to linear combinations of two-qubit Stokes
parameters) for the Werner states. Moreover, we show that the
generalized Werner states (GWSs), which are mixtures of an
arbitrary two-qubit pure state and white noise, can reveal a more
refined hierarchy of the quantum-correlation measures using our
experimental setup.

We note that the setup applied in this work was also used earlier
in Refs.~\cite{Lemr2016, Bartkiewicz2017, Travnicek2021,
Roik2022}, but for conceptually different tasks, e.g., measuring
collective nonlinear witnesses of entanglement~\cite{Rudnicki2011,
Rudnicki2012}, Bell nonlocality measure~\cite{Bartkiewicz2018}, or
diagnosing an entanglement-swapping protocol. The setup enables
entanglement swapping and measuring multicopy entanglement
witnesses as inspired by Refs.~\cite{Horodecki2003,Bovino2005}.

The setup also bears some similarities with a previously proposed
and implemented scheme by Bovino {\it et al.} \cite{Bovino2005}.
Our experimental method of measuring $R$ for general two qubit
states is conceptually similar to that reported in
Ref.~\cite{Bovino2005} for measuring a nonlinear entropic witness.
We find that the witness, defined in the next section, can
actually be interpreted as the three-measurement steering measure
$S$. However, the advantage of our method is that it is more
versatile. As shown in Ref.~\cite{Bartkiewicz2017}, one can
perform a full tomography of all the elements of the $R$ matrix
rather than only determining its trace. Thus, from the set of six
numbers (determining a correlation matrix $R$) we can learn much
more about quantum correlations compared to the original method of
Ref.~\cite{Bovino2005}. In addition to that, our design provides
several practical benefits with respect to Ref.~\cite{Bovino2005}.
Namely from the experimental point of view, it only requires a
single Hong-Ou-Mandel interferometer instead of two. Moreover, our
design shares the same geometry with the entanglement-swapping
protocol~\cite{Bartkiewicz2017}. As a result, it can be deployed
in future teleportation-based quantum networks to acquire various
entanglement measures of distributed quantum states.

This experimental method of measuring the $R$ matrix enables us a
complete determination of not only steering measures, but also a
fully entangled fraction (FEF)~\cite{Bennett1996} and Bell
nonlocality measures~\cite{Horodecki1995}. We note that for the
GWSs, the FEF is exactly equal to the two most popular measures of
entanglement, i.e., the negativity and
concurrence~\cite{HorodeckiReview}. Thus, the hierarchy of the
three measures can be experimentally determined from the $R$
matrix for the Werner states, which is the main goal of this
paper.

\section*{Correlation matrix $R$ for Werner and Werner-like states}
\label{Sec-R}

We study quantum effects in two qubits by means of the $3\times3$
correlation matrix $R=T^T T$, which is defined by the matrix $T$
composed of the two-qubit Stokes parameters
$T_{ij}=\mathrm{Tr}[\rho(\sigma_{i}\otimes\sigma_{j})]$, which are
the mean values of the Pauli matrices $\sigma_{i}$ ($i=1,2,3$).
Superscript $T$ denotes transposition. The standard Bloch
representation of a general two-qubit state $\rho$ can be given by
the elements $T_{ij}$ together with the single-qubit Stokes
parameters $u_{i}=\mathrm{Tr[}\rho(\sigma_{i}\otimes I_{2})]$ and
$v_{i}=\mathrm{Tr[}\rho(I_{2}\otimes\sigma_{i})]$ as
\begin{equation}
\rho  =  \frac{1}{4}\Big(
I_{4}+\boldsymbol{u}\cdot\boldsymbol{\sigma}\otimes
I_{2}+I_{2}\otimes\boldsymbol{v}\cdot\boldsymbol{\sigma}+\!\!\!\sum
\limits_{i,j=1}^{3}T_{ij}\,\sigma _{i}\otimes \sigma _{j}\Big),
\label{rhoGeneral}
\end{equation}
where $\boldsymbol{u}=[u_{1},u_{2},u_{3}]$ and
$\boldsymbol{v}=[v_{1},v_{2},v_{3}]$ denote the Bloch vectors of
the first and second qubits, respectively. Moreover,
$\boldsymbol{\sigma}=[\sigma_{1},\sigma_{2},\sigma_{3}]\equiv
[X,Y,Z]$, and $I_{n}$ is the $n$-qubit identity operator.

We analyze in detail a special type of the general states given in
Eq.~(\ref{rhoGeneral}). Specifically, we have experimentally
generated the polarization Werner states by mixing the singlet
Bell state, $\ket{\psi^{-}}=(\ket{HV}-\ket{VH})/\sqrt{2}$, with
white noise (i.e., the maximally mixed state)~\cite{Werner1989}:
\begin{eqnarray}
  \rho_{\rm W} &=& p\ket{\psi^{-}}\bra{\psi^{-}}+\frac{1-p}{4} I_{4},
\label{WernerState}
\end{eqnarray}
assuming various values of the mixing (noise) parameter
$p\in[0,1]$. Here, $\ket{H}$ and $\ket{V}$ denote horizontal and
vertical polarization states, respectively. The correlation matrix
$R$ for the Werner states simplifies to $R(\rho_{\rm W}) = p^2
I_3$.

We also theoretically analyze GWSs, which can be defined by
replacing the singlet state $\ket{\psi^{-}}$ in
Eq.~(\ref{WernerState}) by a pure state
$\ket{\psi_q}=\sqrt{q}\ket{HV}-\sqrt{1-q}\ket{VH}$ with a
superposition parameter $q\in[0,1]$, i.e.,
\begin{eqnarray}
  \rho_{\rm GW}(p,q) &=& p\ket{\psi_q}\bra{\psi_q}+\frac{1-p}{4}
  I_{4}.
  \label{GWS}
\end{eqnarray}
The state can also be obtained by transmitting a photon in the
state $\ket{\psi_q}$ through a depolarizing channel. Note that
GWSs also often defined slightly differently, i.e., via
$\ket{\phi_q} =\sqrt{q}\ket{HH} +\sqrt{1-q}\ket{VV}$ instead of
$\ket{\psi_q}$ in Eq.~(\ref{GWS}), as experimentally studied in,
e.g., Ref.~\cite{Jirakova2021}. A special case of such states,
i.e., a modified Werner state, when $\ket{\psi^{-}}$ is replaced
by $\ket{\phi_{q=1/2}}$, is referred to as an isotropic state.
Such modifications of the Werner states or the GWSs do not affect
their quantum correlation measures.

The correlation matrix $R$ for the GWSs, given in Eq.~(\ref{GWS}),
is diagonal and reads
\begin{equation}
R[\rho_{\rm GW}(p,q)] =
\begin{pmatrix}
4p^2q(1-q) & 0 & 0 \\
0 & 4p^2q(1-q) & 0 \\
0 & 0 & p^2
\end{pmatrix}.
\label{R_GWS}
\end{equation}
We note that the correlation matrices $T$ and $R$ are in general
nondiagonal (including the non-perfect Werner state measured by us
experimentally), although they are diagonal for the perfect GWSs
states given in Eq.~(\ref{GWS}). Anyway, as shown in
Ref.~\cite{Luo2008}, an arbitrary state $\rho$ described by a
nondiagonal $T$, can be transformed (via a singular-value
decomposition) into a state with a diagonal $T$ by local unitary
operations, thus, without changing its quantum correlations,
including those studied below.

\section*{Measures of quantum correlations for Werner and Werner-like states}
\label{Sec-measures}

\subsection*{Fully entangled fraction and entanglement measures}
\label{Sec-FEF}

The FEF~\cite{Bennett1996} for an arbitrary two-qubit state $\rho$
in Eq.~(\ref{rhoGeneral}) can be defined
as~\cite{Bartkiewicz2017}:
\begin{equation}
   \FEF(\rho) = \tfrac12\theta(\tr \sqrt{R}-1),
  \label{FEF}
\end{equation}
given in terms the function $\theta(x)=\max(x,0)$. In general, the
FEF is only a witness of entanglement; however, for some special
classes of two-qubit states, including the GWSs, the FEF becomes a
good entanglement measure, and it reduces to the concurrence and
negativity:
\begin{eqnarray}
  \FEF[\rho_{\rm GW}(p,q)] &=& N(\rho_{\rm GW}) = C(\rho_{\rm GW}) \nonumber
\\
  &=& \tfrac12\theta\left\{p [1+4 \sqrt{q(1-q)}]-1\right\}.
\label{E_GWS}
\end{eqnarray}
For completeness, we recall that the concurrence $C(\rho)$ of an
arbitrary two-qubit state $\rho$ is defined as~\cite{Hill1997}:
$C(\rho)=\theta(\sqrt{\lambda_1}-\sqrt{\lambda_2}-\sqrt{\lambda_3}-\sqrt{\lambda_4})$,
where $\lambda_1 \geqslant \lambda_2 \geqslant \lambda_3 \geqslant
\lambda_4$ are the eigenvalues of $\rho(\sigma_2 \otimes
\sigma_2)\rho^*(\sigma_2 \otimes \sigma_2)$, the superscript $*$
denotes complex conjugation, and $\sigma_2$ is the second Pauli
matrix. Moreover, we recall the definition of the negativity $N$
of a two-qubit state $\rho$, which reads~\cite{Zyczkowski1998}:
${N}({\rho})=\theta(-2\mu _{\min})$ with $\mu_{\min}$ denoting the
smallest eigenvalue of $\rho^{\Gamma}$, i.e., $\min[{\rm
eig}(\rho^{\Gamma})]$, where the superscript $\Gamma$ indicates
partial transposition. It is seen that the negativity,
concurrence, and FEF reduce to the same function for the GWSs.

Let $p_E(q)$ denote the largest value of the mixing parameter $p$
as a function of the superposition parameter $q$ for which
$\rho_{\rm GW}(p,q)$ is separable. This can be obtained by solving
$\FEF(\rho_{\rm GW})=0$ resulting in:
\begin{equation}
    p_E(q)=1/\big[1+4 \sqrt{q(1-q)}\big],
  \label{p_E}
\end{equation}
which means that $\rho_{\rm GW}(p,q)$ is entangled iff
$p\in(p_E(q),1]$. In the special case of the standard Werner
states, Eq.~(\ref{E_GWS}) simplifies to
\begin{equation}
  \FEF[\rho_{\rm W}(p)]=N[\rho_{\rm W}(p)]= C[\rho_{\rm W}(p)] =
  \theta(3p-1)/2,
\label{E_WS}
\end{equation}
which implies the well known fact~\cite{Werner1989} that the
Werner state is separable iff the mixing parameter $p\in[0,1/3]$.

It should be noted that entanglement measures for general states,
given in Eq.~(\ref{rhoGeneral}), depend not only on the
correlation matric $R$, but also on the single-qubit Stokes
parameters $\langle \sigma_n^{i}\rangle$ for $n=1,2,3$ and
$i=1,2$. It is seen that the FEF is not a universal witness of
two-qubit entanglement, because it solely depends on the $R$
matrix. Nevertheless, the FEF is a good measure of the
entanglement of the GWSs.

\subsection*{Quantum steering measures}
\label{Sec-S}

The effect of quantum steering of a two-qubit state refers to the
possibility to affect at a distance one qubit (say of Bob) via
local measurements performed on the other qubit (say by Alice).
The quantum steerability of a given two-qubit state $\rho$ can be
experimentally tested, assuming that each party is allowed to
measure $n$ observables in their sites (qubit), by the inequality
derived by Cavalcanti, Jones, Wiseman, and Reid (CJWR), which
reads~\cite{Cavalcanti2009}:
\begin{equation}
  F_n(\rho,{\bf r}) =\frac{1}{\sqrt{n}}
\left|\sum_{i=1}^n\langle A_i\otimes B_i\rangle\right| \leqslant 1,
  \label{CJWR}
\end{equation}
where ${\bf r} =\{\hat{r}^A_1,...,\hat{r}^A_n, \hat{r}^B_1, ...,
\hat{r}^B_n\}$ is the set of measurement directions with
$\hat{r}^A_i,\hat{r}^B_i\in\mathbb{R}^3$ (for $i=1,...,n$)
denoting unit and orthonormal vectors, respectively. Moreover,
$A_i = \hat{r}^A_i\cdot\boldsymbol{\sigma}$, $B_i =
\hat{r}^B_i\cdot\boldsymbol{\sigma}$,  and $\langle A_i\otimes
B_i\rangle=\text{Tr}(\rho A_i\otimes B_i)$. A measure of steering
can be obtained by maximizing $F_n(\rho,{\bf r})$ over the set of
measurement directions, i.e., $F_n(\rho)=\max_{\bf r}F_n(\rho,{\bf
r})$. More specifically, Costa and Angelo~\cite{Costa2016}
suggested the following steering measures depending on the number
$n$ of measurements per qubit:
\begin{equation}
  S_n(\rho)={\cal N}_n \theta[F_n(\rho)-1],
  \label{Sn}
\end{equation}
where ${\cal N}_n=[\max_{\rho} F_n(\rho)-1]^{-1}$ is the
normalization constant such that $S_n(\rho)\in[0,1]$ for any
two-qubit $\rho$. Hereafter, we focus on analyzing the steering
measures $S_2$ and $S_3$ (and related quantifiers) in the two- and
three- measurement scenarios, denoted as 2MS and 3MS, which
correspond respectfully to measuring two and three Pauli operators
on qubits of both parties. Costa and Angelo found that these
two-qubit steering measures can be compactly written
as~\cite{Costa2016}:
\begin{equation}
  S_3 (\rho)= \frac{\theta(c-1)}{\sqrt{3}-1},\quad
  S_2 (\rho)= \frac{\theta(\sqrt{c^2-c^2_{\min}}-1)}{\sqrt{2}-1},
\label{S23}
\end{equation}
respectively, given in terms of $c=\sqrt{c_1^2+c_2^2+c_3^2}$ and
$c_{\min}=\min |c_i|$, where $\{c_i\}={\rm svd}(T)$ are singular
values of $T$. Note that the original formulas for $S_2$ and $S_3$
in Ref.~\cite{Costa2016} were given assuming the diagonal form of
the matrix $T$, so $c_i$ were simply given by $T_{ii}$. The
steering measures given in~(\ref{S23}) can be rewritten in terms
of the correlation matrix $R$ as follows:
\begin{eqnarray}
  S_3(\rho) &=& \frac{\theta(\sqrt{\tr R}-1)}{\sqrt{3}-1},\label{S3}\\
  S_2(\rho) &=& \frac{\theta\big\{\sqrt{\tr R-\min[{\rm eig}(R)]}-1\big\}}{\sqrt{2}-1},
\label{S2}
\end{eqnarray}
The Costa-Angelo measure $S_3$ of steering in the 3MS is sometimes
modified as~(see, e.g., Refs.~\cite{Fan2021,Yang2021}):
\begin{equation}
  S (\rho) = \sqrt{\tfrac 12\theta(\tr R-1)},
  \label{steeringS}
\end{equation}
and we also apply this measure in the following, because of a
useful property that $S$ reduces to the negativity and concurrence
for any two-qubit pure states. Note that $S,S_3\in [0,1]$ and they
are monotonically related to each other for any two-qubit states:
\begin{equation}
  S_3(\rho)=\frac{\sqrt{2S^2(\rho)+1}-1}{\sqrt{3}-1}\le S(\rho).
  \label{SS3}
\end{equation}
For the GWSs, described by the correlation matrix $R$ given in
Eq.~(\ref{R_GWS}), we find
\begin{eqnarray}
  S[\rho_{\rm GW}(p,q)] &=&
  \sqrt{\tfrac12\theta[8p^2q(1-q)+p^2-1]},
  \label{S_GW}
  \\
  S_3[\rho_{\rm GW}(p,q)] &=& \frac{\theta[p\sqrt{1+8q(1-q)}-1]}{\sqrt{3}-1}.
  \label{S3_GW}
\end{eqnarray}
Let $p_S(q)$ denote the largest value of the mixing parameter $p$
for a given value of the superposition parameter $q$ for which
$\rho_{\rm GW}(p,q)$ is unsteerable. Thus, by solving $S(\rho_{\rm
GW}) =0$, we have:
\begin{equation}
  p_S(q)=[1+8q(1-q)]^{-1/2},
  \label{p_S}
\end{equation}
which means that a given GWS, $\rho_{\rm GW}(p)$, is steerable
assuming three measurements per qubit iff the mixing parameter
$p\in(p_S(q),1]$. In the special case of the Werner states,
Eq.~(\ref{S_GW}) simplifies to the formulas:
\begin{equation}
  S[\rho_{\rm W}(p)] = \sqrt{\tfrac12\theta(3p^2-1)},\quad
  S_3[\rho_{\rm W}(p)] = \frac{\theta(\sqrt{3}p-1)}{\sqrt{3}-1},
  \label{S_W}
\end{equation}
which imply that $\rho_{\rm W}(p)$ is unsteerable in the 3MS iff
$p\in [0,1/\sqrt{3}]$.

Quantum steerability in the 2MS, as based on $S_2$ or related
measures, corresponds to Bell nonlocality and it is discussed in
detail in the next section.

We note that to quantify steering, assuming three measurements on
both Alice' and Bob's qubits, we can interchangeably use $S_3$,
defined in Eq.~(\ref{S3}), $S$ in Eq.~(\ref{steeringS}), as well
the steerable weight~\cite{Skrzypczyk2014} (as applied in our
closely related paper~\cite{Jirakova2021}), or the steering
robustness~\cite{Piani2015} in the 3MS. Indeed, if one of the
steering measures vanishes, then all the other measures vanish
too. However, the steering measure $S_2$, as defined in
Eq.~(\ref{S2}) in the 2MS, although it is equivalent to the Bell
nonlocality measure $B$, but it is fundamentally different from
another steering measure $S_2$ (for clarity denoted here as
$S'_2$) studied by us in Ref.~\cite{Jirakova2021}, because it
corresponds to the case when Alice (Bob) performs two (three)
measurements on her (his) qubit. Thus, $S_2(\rho)=0$
(corresponding to vanishing Bell nonlocality of a given state
$\rho$) does not imply that also $S'_2(\rho)=0$, which was shown
experimentally in~\cite{Jirakova2021}). This is possible because
an extra measurement performed by Bob on his qubit, as allowed in
the $S_2'$ scenario, can reveal the steerability of $\rho$.

\subsection*{Bell nonlocality measures}
\label{Sec-B}

The Bell nonlocality of a given two-qubit state $\rho$ can be
tested by the violation of the Bell inequality in the
Clauser-Horne-Shimony-Holt (CHSH) form \cite{Clauser1969}
\begin{equation}
|\langle {\cal
B}\rangle_{\rho}|\equiv\big|\big\langle\boldsymbol{a}\cdot
\boldsymbol{\sigma }\otimes (\boldsymbol{
b}+\boldsymbol{b}^{\prime })\cdot \boldsymbol{\sigma
}+\boldsymbol{a}^{\prime }\cdot \boldsymbol{\sigma }\otimes
(\boldsymbol{b}-\boldsymbol{b}^{\prime })\cdot \boldsymbol{\sigma
}\big\rangle_{\rho}\big|\leq 2, \label{CHSH}
\end{equation}
where $\boldsymbol{a}, \boldsymbol{a'}, \boldsymbol{b},
\boldsymbol{b'}\in\mathbb{R}^3$ are unit vectors describing
measurement settings, and ${\cal B}$ is referred to as the
Bell-CHSH operator. Bell nonlocality can be quantified by the
maximum possible violation of the CHSH inequality in~(\ref{CHSH})
over all measurement settings, which lead Horodecki \emph{et al.}
to the following analytical formula \cite{Horodecki1995}
\begin{equation}
\max_{\nu}\langle {\cal B}\rangle_{\rho}=2\sqrt{M(\rho)},
\label{CHSH_max}
\end{equation}
where the nonnegative quantity $M(\rho)$ is the sum of the two
largest eigenvalues of $R(\rho)$. The CHSH inequality in
(\ref{CHSH}) is satisfied iff $M(\sigma)\le 1$. For a better
comparison with other measures of quantum correlations defined in
the range [0,1], the Bell nonlocality measure of Horodecki
\emph{et al.}~\cite{Horodecki1995} can be given by (see,
e.g.,~\cite{Adam2004,Bartkiewicz2017,Yang2021})
\begin{equation}
  B(\rho) = \sqrt{\theta[M-1]}
  =\sqrt{\theta\big\{\tr R -\min[{\rm eig}(R)]-1\big\}},
\label{nonlocalityB}
\end{equation}
or, equivalently, as~\cite{Costa2016}
\begin{eqnarray}
  B'(\rho)&=&\frac{\theta[\sqrt{M}-1]}{\sqrt{2}-1}
  = \frac{\theta\big(\sqrt{\tr R-\min[{\rm eig}(R)]}-1\big)}{\sqrt{2}-1}
  \nonumber\\
   &=& S_2(\rho),
  \label{nonlocalityBCosta}
\end{eqnarray}
which guarantee that $B,B'\in[0,1]$. It is seen that $B'$ is
exactly equal to the steering measure $S_2$, given in
Eq.~(\ref{S2}), in the 2MS.

Hereafter, we apply both nonlocality measures because their
specific advantages. In particular, as shown explicitly below,
$B'$ depends linearly on the mixing parameter $p$ of the Werner
states and GWSs, thus its experimental estimation results in
smaller error bars compared to those of $B$. On the other hand,
$B$ is equal to the negativity and
concurrence~\cite{Adam2004,Adam2004b}, but also to the steering
measure $S$ and the FEF:
\begin{eqnarray}
   B(\ket{\psi})&=&S(\ket{\psi})=\FEF(\ket{\psi})
   \nonumber \\
   &=&C(\ket{\psi})=N(\ket{\psi})=2|ad-bc|,
  \label{BCN}
\end{eqnarray}
for an arbitrary two-qubit pure state
$\ket{\psi}=a\ket{HH}+b\ket{HV}+c\ket{VH}+d\ket{VV}$, where
$a,b,c,d$ are the normalized complex amplitudes. This useful
property of $B$ is not satisfied for $B'$. We also study $B$ to
enable a more explicit comparison of our present experimental
results with those in our former closely related
papers~\cite{Jirakova2021, Bartkiewicz2017}. Anyway, $B$ and $B'$
are monotonically related to each other:
\begin{equation}
  B'(\rho)=\frac{\sqrt{B^2(\rho)+1}-1}{\sqrt{2}-1} \le B(\rho).
  \label{BvsB}
\end{equation}

The Bell nonlocality measures $B$ and $B'$ for the Werner states
read
\begin{equation}
  B[\rho_{{\rm W}}(p)] = \sqrt{\theta(2p^2-1)},\quad
  B'[\rho_{{\rm W}}(p)] = \frac{\theta(\sqrt{2}p-1)}{\sqrt{2}-1},
  \label{B_WS}
\end{equation}
which explicitly shows that the states are nonlocal iff
$p>1/\sqrt{2}$. By comparing Eq.~(\ref{B_WS}) with
Eq.~(\ref{E_WS}), it is clearly seen that the Werner states for
the mixing parameter $p\in (1/3,1/\sqrt{2})$ are entangled, but
without violating the CHSH inequality, as first predicted in
Ref.~\cite{Werner1989}. For the GWSs, formulas in Eq.~(\ref{B_WS})
generalize to:
\begin{eqnarray}
  B[\rho_{{\rm GW}}(p,q)]  &=& \sqrt{\theta\left\{p^2 [1+4
  q(1-q)]-1\right\}},
    \\
  B'[\rho_{{\rm GW}}(p,q)]  &=& \frac{\theta[p\sqrt{1+4
  q(1-q)}-1]}{\sqrt{2}-1},
\label{B_GWS}
\end{eqnarray}
Let $p_B(q)$ denote the largest value of the mixing parameter $p$
for a given value the superposition parameter $q$ for which
$\rho_{\rm GW}(p,q)$ is Bell local. Thus, by solving $B(\rho_{\rm
GW})=0$ one finds:
\begin{equation}
  p_B(q)=[1+4q(1-q)]^{-1/2},
  \label{p_B}
\end{equation}
which means that $\rho_{\rm GW}(p,q)$ is Bell nonlocal if
$p\in(p_B(q),1]$. This function implies for $q=1/2$ the well-known
result that the Werner state violates the CHSH inequality iff the
mixing parameter $p\in(1/\sqrt{2},1]$~\cite{Werner1989}.

\begin{figure}
\includegraphics[width=\columnwidth]{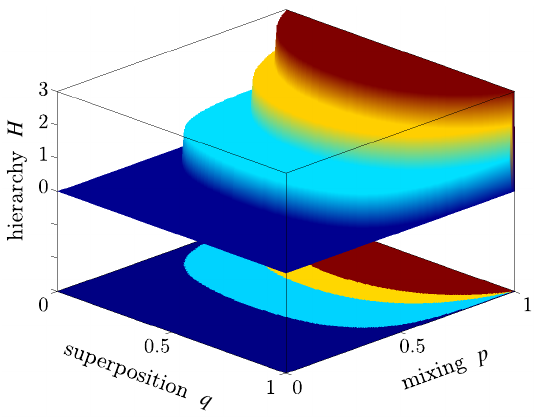}
\caption{Hierarchy of quantum correlations of the generalized
Werner states: The hierarchy parameter $H[\rho_{\rm GW}(p,q)]$,
defined in Eq.~(\ref{H_GWS}), versus the superposition ($q$) and
mixing ($p$) parameters. A given GWS, $\rho_{\rm GW}(p,q)$, is
separable if $H(\rho_{\rm GW})=0$, entangled if $H(\rho_{\rm
GW})\ge 1$, steerable in the 3MS if $H(\rho_{\rm GW})\ge 2$, and
Bell nonlocal (and steerable in the 2MS) if $H(\rho_{\rm GW})=3$.}
\label{fig1}
\end{figure}
\subsection*{Hierarchy of quantum correlations}
\label{Sec-Hierarchy}

The following hierarchy of the discussed quantum correlation
measures hold for a general two-qubit state $\rho$:
\begin{eqnarray}
  B(\rho) \le S(\rho) \le \FEF(\rho) \le N(\rho) \le C (\rho),
\label{ineqs}
\end{eqnarray}
or, equivalently,
\begin{eqnarray}
  S_2(\rho) \le S_3(\rho) \le \FEF(\rho) \le N(\rho) \le C (\rho),
\label{ineqs}
\end{eqnarray}
We also note that $S_2(\rho) \le B(\rho)$ and $S_3(\rho) \le
S(\rho)$. The inequalities in~(\ref{ineqs}) for the GWSs reduce to
\begin{equation}
  B(\rho_{\rm GW}) \le S(\rho_{\rm GW}) \le \FEF(\rho_{\rm GW}) = N(\rho_{\rm GW}) =
  C(\rho_{\rm GW}).
\label{ineqs_werner}
\end{equation}
To visualize this hierarchy, we define the following hierarchy
parameter of quantum correlations for the GWSs,

\begin{eqnarray}
  H(\rho_{\rm GW})=
   \chi[B(\rho_{\rm GW})]+
   \chi[S(\rho_{\rm GW})]+
   \chi[\FEF(\rho_{\rm GW})]
   \nonumber \\
     = \chi[S_{2}(\rho_{\rm GW})]+
   \chi[S_{3}(\rho_{\rm GW})]+
   \chi[\FEF(\rho_{\rm GW})],\quad\quad
  \label{H_GWS}
\end{eqnarray}
which is given in terms of the Heaviside function $\chi(x)$ equal
to 1 for $x>0$ and zero for $x\le 0$. This parameter is plotted in
Fig.~\ref{fig1} as a function of the parameters $p$ and $q$
uniquely specifying $\rho_{\rm GW}(p,q)$.

\section*{Experimental setup}
\label{Sec-setup}

\begin{figure}
\includegraphics[width=\columnwidth]{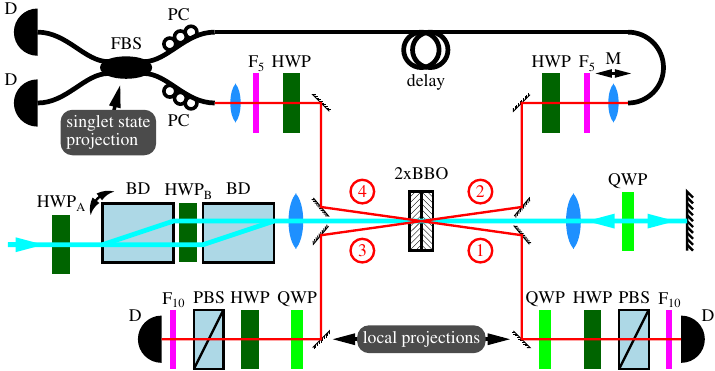}
\caption{\label{fig2} Schematic depiction of the experimental
setup. Individual components are labelled as follows: HWP --
half-wave plate, QWP -- quarter-wave plate, D -- detector, BD --
beam displacer, PBS -- polarizing beam splitter, BBO --
$\beta$-barium-borate crystals, M -- motorized translation,
F$_{5,10}$ -- 5, 10~nm-wide bandpass filters, FBS -- fiber beam
splitter, PC -- polarization controller. The photons generated
during the forward (backward) propagation of pump photons through
the BBO crystals are labelled as 1 and 2 (3 and 4).}
\end{figure}

The experiment is implemented on the platform of linear optics
with qubits encoded into polarization states of discrete photons.
These photons are generated in the process of spontaneous
parametric down-conversion occurring in a cascade of two Type-I
BBO crystals in the Kwiat \emph{et al.}
configuration~\cite{Kwiat1995}. A femtosecond fundamental laser
pulse is frequency doubled to 413~nm and pumps the crystal cascade
on its way there and back (as depicted in Fig.~\ref{fig2}). Each
time the pulse impinges on the crystals, a pair of photons can be
generated in the polarization singlet Bell state. To achieve a
high degree of entanglement, the pumping pulse is diagonally
polarized (by the half-wave plate HWP$_\mathrm{A}$) and subject to
a polarization dispersion line~\cite{Nambu2002}. In our case, this
dispersion line is implemented by two beam displacers (BDs)
enveloping HWP$_\mathrm{B}$. The photons generated, while the
pulse propagates forward are labelled 1 and 2 while the photons
generated in the pulses second-time travel through the crystals
are denoted 3 and 4.

The investigated state is encoded both into photons 1 and 2 (the
first copy) and into photons 3 and 4 (the second copy). A
collective measurement on both copies is then performed by
projecting photons 2 and 4 onto the singlet Bell state using a
fiber beam splitter (FBS) followed by post-selection onto
coincidence detection on its output ports. The remaining photons 1
and 3 are projected locally by means of the sets of quarter and
half-wave plates (QWPs and HWPs) and polarizing beam splitters
(PBSs). We recorded the number of four-fold coincidence detections
for various settings of the wave plates; namely, for all
combinations of the projections onto the horizontal, vertical,
diagonal, anti-diagonal, and both circular polarization states. We
have subsequently calculated the expectation values of the Pauli
matrices, that is $A_{ij}=\tr[\rho_1\rho_2\Pi\sigma_i\sigma_j]$,
where $\Pi=-4\ket{\psi^{-}}\bra{\psi^{-}}$. Note that this formula
is almost identical to the one in our previous
paper~\cite{Bartkiewicz2017}, except that in the paper instead of
the $\Pi$ projection, the $1-\Pi$ projection was applied there.

When adjusting the setup to generate the requested Bell state (or
the maximally mixed state), we have tuned the polarization of the
pump beam, so that locally the generated photons have equal
probabilities to be horizontally and vertically polarized. (The
probability for a single photon being horizontally polarized is
$p_H=0.50\pm0.03$.) Balancing these probabilities for the
horizontal and vertical polarizations implies also balancing in
any single-photon polarization basis. Note that the single-photon
state is fully incoherent, because the other photon from a pair is
ignored and, hence, mathematically one traces over its state. As a
consequence, we can consider $B_{ij}=\tr[\rho_1\rho_2
I_{4}\sigma_i\sigma_j] \approx 0$. With respect to that we
conclude that the prepared two copies of the Bell state are
balanced enough to warrant the replacement of $1-\Pi$ by $\Pi$.
This is also supported by the fact that the numerically closest
Bell state producing the observed values for the three measures
has its parameter $q=0.474$ -- see Eq.~(\ref{eq:dephasedBell}) and
comments in the surrounding paragraph.

Despite narrow frequency filtering on all photons (see the
parameters of the bandpass filters in Fig.~\ref{fig2}) and a
relatively thin crystal cascade of twice 1~mm, there is a
generation-time jitter, which causes the visibility of two-photon
interference on the FBS to decrease. We have performed a
calibration measurement that reveals that 56.7\% of the photons do
not interfere on FBS. Moreover, the laser power fluctuates over
time yielding variable rates of photon-pairs generation. In order
to compensate for these two effects, we have performed all the
measurements in the two regimes with a temporal delay between
photons 2 and 4: (a) tuned for interference and (b) detuned
(controlled by the motorized translation M). These two
measurements together with the calibration measurement allow us to
estimate the net probability of the two copies of the investigated
state to pass simultaneously the Bell-state projection on the FBS,
as well as the local polarization projections resulting in a
four-fold detection event. With the repetition rate of the laser
pulse of 80~MHz, we achieve about 1 such an event per 5 minutes.

While the crystals generate two copies of the singlet Bell state,
we can readily modify the detection electronics to effectively
perform the measurement on the two copies of a maximally mixed
state. So far the coincidence window, i.e., the time within all
photons must be detected to be considered a coincidence event, had
to be very narrow (5~ns) to assure detection of photon pairs
originating from a single laser pulse. By considerably widening
that window by several orders of magnitude, we effectively
aggregate also detections that are completely unrelated and
mutually random. This way, the observed state becomes effectively
white noise.

Having all the measurements performed on a pure entangled state
(two copies of the singlet Bell states) as well as on the
maximally mixed state (i.e., the two copies of the maximally mixed
state), we can easily interpolate the results for any Werner state
with mixing parameter $p$. In order to do so, we make use of the
fact that when two polarization states of single photons interact
on a beam splitter and one of them is being a maximally mixed
state, the resulting probability of coincidence detection is
independent of the state of the other photon. As a result, we
interpolate the measurement for any Werner state by combining with
probability $p^2$ the outcomes observed on two copies of maximally
entangled states and with probability of $1-p^2$ the results
observed on a maximally mixed state.

Note that, contrary to reconstructing the $R$ matrix,  there is no
experimental advantage of reconstructing the $3\times 3$ matrix
$T\equiv T_3$ compared to a full QST of a two qubit state $\rho$,
which corresponds to reconstructing the $4\times 4$ matrix
$T_4=[\<\sigma_n\otimes \sigma_m\>]$ for $n,m=0,...,3$, where
$\sigma_0=I_2$ is the qubit identity operator. It might look that
reconstructing all the 9 elements of $T_3$ is much simpler than
reconstructing 16 (or 15) elements of $T_4$. But this is not the
case, because the required types of measurements are the same in
both reconstructions. Note that the optical reconstruction $T_3$
for a given two-qubit polarization state $\rho$ is usually based
on projecting $\rho$ on all the eigenstates of the three Pauli
operators  for each qubit, i.e., projections onto the six
polarization single-qubit states (so 36 two-qubit states):
diagonal ($|D\>$), antidiagonal ($|A\>$), right- ($|R\>$) and
left-circular ($|L\>$), horizontal ($|H\>$), and vertical
($|V\>$). Analogously, a standard QST of $\rho$ also corresponds
to reconstructing $T_4$ via the same 36 projections as those for
$T_3$, and the single-qubit identity operator is given by
$I_{2}=|H\>\<H|+|V\>\<V|$. So, the required measurements for
reconstructing $T_3$ and $T_4$ are the same, but only their
numerical reconstructions are different, although can be based on
exactly the same measured data.

\section*{Experimental hierarchy of quantum correlations}
\label{Sec-exp-hierarchy}
\def \dd#1#2{\,[{+#2},{-#1}]}

\begin{figure}[]
\centering
\includegraphics[width=\columnwidth]{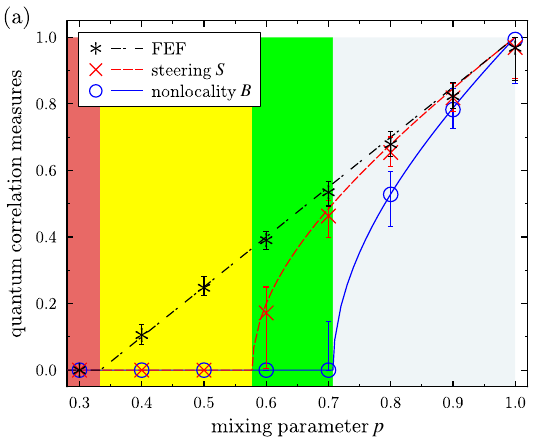}
\includegraphics[width=\columnwidth]{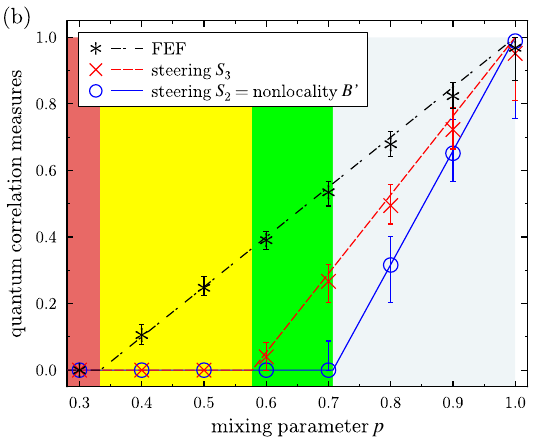}
\caption{Experimental demonstration of the hierarchy of quantum
correlations of the Werner states without full QST: the Bell
nonlocality measures (a) $B$ and (b) $B'=S_2$ (solid blue lines
and curves), the 3MS steering measures (a) $S$ and (b) $S_3$
(dashed red), and (a,b) the FEF (dot-dashed black lines) shown
versus the mixing parameter $p$. Symbols depict experimental
results and curves represent theoretical predictions.
  \label{fig3}}
\end{figure}

\begin{figure}[]
\centering
\includegraphics[width=0.88\columnwidth]{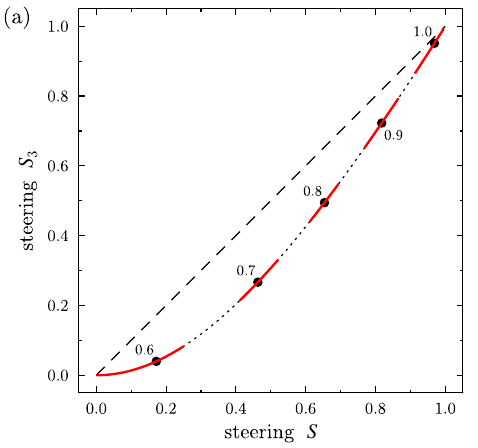}
\includegraphics[width=0.88\columnwidth]{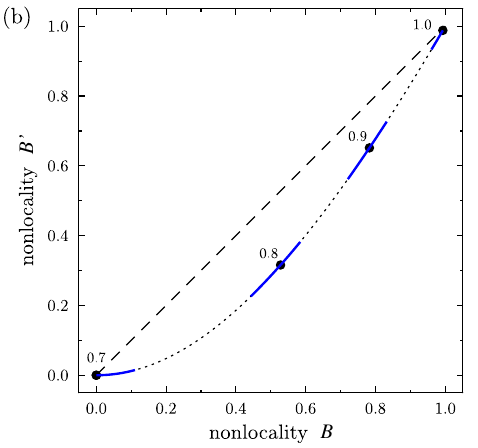}
\caption{Experimental and theoretical predictions of different
measures: (a) $S_3$ vs $S$ quantifying steering in the 3MS and (b)
$S_2=B'$ vs $B$ describing Bell nonlocality and, equivalently,
steering in the 2MS. Symbols depict the measures calculated for
the experimental Werner states for the indicated values of the
mixing parameter $p$. The error bars are marked by solid red
curves that follow the dotted curves. Arbitrary two-qubit states
lie on the dotted curves. The dashed diagonal lines are added just
to show the curvature of the solid curves more clearly and stress
the mutual inequality of the plotted quantities. \label{fig4}}
\end{figure}

\begin{table*}
\caption{Quantum correlation measures for the experimental and
theoretical Werner states plotted in Fig.~\ref{fig3}(a), including
measures of Bell nonlocality ($B$) and steering ($S$) in the 3MS,
and the FEF. Experimental values are listed together with their
asymmetric errors in square brackets.}
  \label{table1}
\begin{ruledtabular}
\begin{tabular}{c c c c c c c}
 & \multicolumn{2}{c}{$B$} & \multicolumn{2}{c}{$S$} & \multicolumn{2}{c}{FEF}\\
\cline{2-3} \cline{4-5} \cline{6-7}
$p$ & theory & experiment & theory & experiment & theory & experiment \\
\hline
0.3 & 0.000 & $0.000$                   & 0.000 & $0.000$                   & 0.000 & $0.000$\\
0.4 & 0.000 & $0.000$                   & 0.000 & $0.000$                   & 0.100 & $0.106 \dd{0.030}{0.031}$\\
0.5 & 0.000 & $0.000$                   & 0.000 & $0.000$                   & 0.250 & $0.248 \dd{0.022}{0.034}$\\
0.6 & 0.000 & $0.000$                   & 0.200 & $0.172 \dd{0.168}{0.078}$ & 0.400 & $0.391 \dd{0.028}{0.027}$\\
0.7 & 0.000 & $0.000 \dd{0.000}{0.145}$ & 0.485 & $0.463 \dd{0.064}{0.046}$ & 0.550 & $0.534 \dd{0.041}{0.032}$\\
0.8 & 0.529 & $0.528 \dd{0.098}{0.068}$ & 0.678 & $0.654 \dd{0.043}{0.047}$ & 0.700 & $0.679 \dd{0.038}{0.038}$\\
0.9 & 0.787 & $0.783 \dd{0.057}{0.064}$ & 0.846 & $0.818 \dd{0.041}{0.045}$ & 0.850 & $0.824 \dd{0.038}{0.042}$\\
1.0 & 1.000 & $0.993 \dd{0.133}{0.007}$ & 1.000 & $0.969 \dd{0.092}{0.030}$ & 1.000 & $0.969 \dd{0.098}{0.031}$\\
\end{tabular}
\end{ruledtabular}
\end{table*}


\begin{table*}
\caption{The Costa-Angelo measures $S_3$ and $S_2=B'$ of steering
in the 3MS and 2MS, respectively, for the experimental and
theoretical Werner states plotted in Fig.~\ref{fig3}(b).
  \label{table2}}
\begin{ruledtabular}
\begin{tabular}{c c c c c}
 & \multicolumn{2}{c}{$S_2$} & \multicolumn{2}{c}{$S_3$}\\
\cline{2-3} \cline{4-5}
$p$ & theory & experiment & theory & experiment \\
\hline
0.3 & 0.000 & $0.000$                   & 0.000 & $0.000$           \\
0.4 & 0.000 & $0.000$                   & 0.000 & $0.000$           \\
0.5 & 0.000 & $0.000$                   & 0.000 & $0.000$           \\
0.6 & 0.000 & $0.000$                   & 0.054 & $0.040 \dd{0.040}{0.044}$ \\
0.7 & 0.000 & $0.000$                   & 0.290 & $0.267 \dd{0.064}{0.050}$ \\
0.8 & 0.317 & $0.316 \dd{0.112}{0.086}$ & 0.527 & $0.494 \dd{0.055}{0.062}$ \\
0.9 & 0.659 & $0.652 \dd{0.085}{0.101}$ & 0.763 & $0.723 \dd{0.058}{0.066}$ \\
1.0 & 1.000 & $0.989 \dd{0.233}{0.011}$ & 1.000 & $0.952 \dd{0.141}{0.048}$ \\
\end{tabular}
\end{ruledtabular}
\end{table*}

In this section we test the experimental Werner states generated
in the setup described in the former section and compare
experimental results with theoretical predictions for ideal Werner
states. One can calculate the correlation matrix elements $R_{ij}$
following the derivations~\cite{Bartkiewicz2017}:
\begin{equation}
  R_{ij} = A_{ij} + B_{ij} = {\rm Tr}(\rho_1\rho_2\Pi\sigma_i\sigma_j)
                 + {\rm Tr}(\rho_1\rho_2I_4 \sigma_i\sigma_j),
\end{equation}
noting that $A_{ij}$ and $B_{ij}$ can be experimentally
determined. As a result the physical correlation matrices $R_{ij}$
of the Bell singlet state and the maximally mixed state were
obtained using a maximum likelihood method. First we derive the
correlation matrix $R_{\bs}$ for the singlet Bell state.
\begin{equation}
R_{\bs} = \left(
  \begin{array}{ccc}
   0.971 & 0.073 & 0.010\\
   0.073 & 0.966 &-0.009\\
   0.010 &-0.009 & 0.941\\
  \end{array}
 \right).
 \label{eq:RBell}
\end{equation}
Then we evaluated also the correlation matrix for the maximally
mixed state corresponding to white noise,
\begin{equation}
R_{I} = \left(
  \begin{array}{ccc}
   0.017 & 0.006 & -0.007\\
   0.006 & 0.013 &  0.016\\
   -0.007 & 0.016 & 0.006\\
  \end{array}
 \right).
 \label{eq:Rnoise}
\end{equation}
%
Using definition (\ref{WernerState}) we can derive the correlation
matrix $R_{\rm W}(p)$ of the Werner states for selected values of
the mixing parameter $p$ as follows,
\begin{equation}
  R_{\rm W}(p) = p^2 R_{\bs} + (1-p^2) R_{I}.
\end{equation}
Now we apply the above-described definitions of the quantifiers of
quantum correlations including the defined measures of Bell
nonlocality ($B$ and $B'=S_{2}$), steering in the 3MS ($S$ and
$S_3$), and entanglement (FEF) based on this correlation matrix.

Our experimental results are summarized in Tables~\ref{table1}
and~\ref{table2} and plotted as symbols in Figs.~\ref{fig3}
and~\ref{fig4}. The error bars were derived using a Monte Carlo
method following the normal distribution of the correlation matrix
components with variance corresponding to the number of detected
photocounts. 
Asymmetry of estimated error bars results from presence 
of the $\theta$ function in the formulas for estimated quantities
as well as from the requirement on physicality of the $R$ matrices.
Figure~\ref{fig3} shows also the theoretically
predicted correlation measures plotted with solid curves, which
were calculated for the ideal Werner states.

In the theoretical section we considered the Costa-Angelo steering
measures $S_2$ and $S_3$ that can be calculated also from the $R$
matrix. We evaluated these steering measures using Eqs. (\ref{S3})
and (\ref{S2}). The results are plotted in Fig.~\ref{fig3}(b). It
is clear that these measures linearly depend on the mixing
parameter $p$. The nonzero regions of the correlation measures,
shown in both panels of Fig.~\ref{fig3}, are the same.
Experimental results shown in Fig.~\ref{fig3}(b) are also
summarized in Table~\ref{table2}.

The original correlation matrices $R$ were derived from measured
coincidences using two methods of maximum likelihood estimation of
Ref.~\cite{Hradil2004}. Both the methods lead to the $R$ matrices
that are essentially the same. Our experimental results shown in
Fig.~\ref{fig3} demonstrate a very good agreement with theoretical
predictions. It is clear that $S_2$, $S_3$, and FEF are the most
stable measures at least for the Werner states and GWSs by
exhibiting the smallest errors because of their linear dependence
on the mixing parameter $p$. By contrast to those quantifiers, the
measures of steering $S$ in the 3MS and of Bell nonlocality ($B$)
are much steeper functions and that is why they are much more
sensitive to unavoidable fluctuations of measured coincidence
counts, as reflected in all the derived quantities. A comparison
of the steering measures $S_3$ and $S$ and the Bell nonlocality
measures $S_2$ and $B$ for arbitrary theoretical states and the
experimental Werner states are shown in Fig.~\ref{fig4}.

In the experiment all imperfections of individual components
decrease the resulting correlation measures. Together with the
instability and a natural Poisson randomness of the measured
coincidences, these effects result in measurement uncertainties.
Also our experimentally generated singlet-like state is not
perfect. We tried to simulate all these mentioned imperfections
degrading the input Bell-like state assuming the rest of the
measurement to be nearly perfect. These expected imperfections
result in a class of generalized states in the form of
\begin{equation}
\label{eq:dephasedBell}
\rho = p |\psi^-\rangle\langle\psi^-| + (1-p) |\psi^+\rangle\langle\psi^+|,
\end{equation}
where $|\psi^\pm\rangle = \sqrt{q} |HV\rangle \pm \sqrt{1-q}
|VH\rangle$. The correlation measures for our most entangled
experimental Bell-like state read: $B = 0.9933$, $S = 0.9691$, and
FEF = 0.9685.  We found that these results are most consistent
with a state of the type specified in Eq.~(\ref{eq:dephasedBell})
for the parameters $q \approx 0.474$ and $p\approx 0.994$. This
implies the purity of this Bell-like state of about 98.9\%.

\section*{Conclusions}
\label{Sec-discussion}

We reported the detection of quantum correlation measures of two
optical polarization qubits without QST. Specifically, we have
measured all the elements of the correlation matrix $R$ (which is
symmetric by definition) for the Werner states with different
amount of white noise. These elements correspond to linear
combinations of two-qubit Stokes parameters. With the matrix $R$,
we were able to determine various measures of quantum
entanglement, steerability, and Bell nonlocality of the Werner and
Werner-like states.

Most notably, our experiment allows us to show the hierarchy of
the tested quantum correlation measures. This means that iff the
mixing parameter $p\le 1/3$, the Werner state is separable. For $p
\in (1/3, 1/\sqrt{3}]$, the Werner state exhibits entanglement (as
revealed by the FEF), but it is unsteerable and Bell local.
Subsequently, for $p \in (1/\sqrt{3}, 1/\sqrt{2}]$ the state is
entangled and steerable, but without exhibiting Bell nonlocality.
Finally, for $p>1/\sqrt{2}$ the state is also Bell nonlocal. These
regions, separated by the three values of $p = \{1/3, 1/\sqrt{3},
1/\sqrt{2}\} \approx \{0.333, 0.577, 0.707\}$, are depicted with
different background colors in Fig.~\ref{fig3}. We have also
analyzed theoretically a hierarchy (shown in Fig.~\ref{fig1}) of
some measures of quantum correlations for generalized Werner
states, which are defined as arbitrary superpositions of a
two-qubit partially-entangled pure state and white noise.

The problem of detecting measures of quantum correlations is
essential to assess their suitability for quantum-information
protocols especially for quantum communication and cryptography
when considering not only trusted but also untrusted devices. We
believe that experimental determination of various measures of
entanglement, steering, and Bell nonlocality without full QST, as
reported in this work, clearly shows its advantage compared to
standard methods based on a complete QST. Specifically, our method
relies on measuring only 6 real elements instead of 15 (or even
16) elements in a complete two-qubit QST.

Moreover, experimental studies of a hierarchy of
quantum-correlation measures might be useful for effective
estimations of one measure for a specific value of another without
full QST for experimental quantum channels.

{\bf Acknowledgements}\\

S.A., K.B., and A.M. are supported by the Polish National Science
Centre (NCN) under the Maestro Grant No. DEC-2019/34/A/ST2/00081.

{\bf Author contributions}\\
The main idea of the manuscript was conceived by A.M.; S.A. and
J.S. performed numerical and analytical calculations; the concept
of linear-optical implementation was devised by K.B.; the
experiment was designed, performed, and raw-data processed by
A.\v{C}. and K.L.  All the authors were involved in writing and
discussing the manuscript.

{\bf Competing interests}\\
The authors declare no competing interests.

{\bf Data availability}\\
All the data necessary to reproduce the results are included in
this published article and its digital supplement. We note that
all the raw experimental data used in this work were obtained in
our experiment reported in Ref.~\cite{Bartkiewicz2017}. Of course,
their usage and interpretation are very different here compared to
the previous work.

\bibliographystyle{FabrizioStyle}
\bibliography{hierarchy_bib}
\end{document}